\documentclass[twocolumn,showpacs,preprintnumbers,amsmath,amssymb]{revtex4}

\newcommand{\ds}{\displaystyle}

\newcommand{\be}{\begin{equation}}
\newcommand{\en}{\end{equation}}
\newcommand{\bea}{\begin{eqnarray}}
\newcommand{\ena}{\end{eqnarray}}

\usepackage{graphicx}
\usepackage{dcolumn}
\usepackage{bm}

\begin{document}

\title{The curvature of the universe and the dark energy potential}

\author{  Sergio del Campo $^{a}$}
\affiliation{$^a$ Instituto de F\'\i sica, Universidad Cat\'olica de Valpara\'\i so,\\
Casilla 4059, Valpara\'\i so, Chile.}

\date{\today}

\begin{abstract}
 The flatness of an accelerating universe model (characterized
by a dark energy scalar field $\chi$) is mimicked from a curved
model that is filled with, apart from the Cold Dark Matter
component, a quintessence-like scalar field $Q$. In this process
we characterize the original scalar potential $V(Q)$ and the
mimicked scalar potential $V(\chi)$ associated to the scalar
fields $Q$ and $\chi$, respectively. The parameters of the
original model are fixed through the mimicked quantities that we
relate to the present astronomical data, such that the equation
state parameter, $w_{_{\chi}}$ and the dark energy density
parameter, $\Omega_{\chi}$.
\end{abstract}

\maketitle

\section{\label{sec:level1} Introduction}

At the moment, we do not precisely know   the amount of matter
present in the universe, but we do not as yet know what the
geometry of the universe is. Astronomical observations conclude
that the matter density related to baryonic and nonbaryonic cold
dark matter is much less than the critical density
value~\cite{Wh-etal}. However, measurements of anisotropy in the
cosmic microwave background radiation indicate that the total
matter present in the universe is very much the same than its
critical value~\cite{Ma-etal,RoHa,DB-etal,Ha-etal}. These
measurements agree with the theoretic prediction of inflationary
universe scenarios~\cite{AG}, which predicts that our universe
should become flat after leaving the inflationary era. In light of
these results, it seems that there exist an important amount of
matter that we are not considering.

On the other hand, recent measurements of type Ia supernova at
high redshift indicate that our universe is
accelerating~\cite{Ri-etal,Pe-etal}. The most simple description
of this acceleration is that it can be characterized as a
cosmological constant, which contributes to a negative pressure.
Other approaches have been employed for explaining this
acceleration. We distinguish that related to the quintessence
model. This model is characterized by an evolving scalar field,
$\chi$ and its scalar-field potential, $V(\chi)$~\cite{CaDaSt}.

Any cosmological model becomes characterized by the total matter
density parameter, $\Omega_T$. This parameter is defined by the
ratio between the total matter ($\rho_{_{T}}$) and the critical
energy ($\rho_{_{C}}$) densities. For instance, in the case of the
Cosmological Constant ($\Lambda$) Cold Dark Matter (CDM) model,
this parameter is given by $\Omega_T = \Omega_M +
\Omega_{\Lambda}$, in which $ \ds \Omega_M
=\frac{\rho_M^0}{\rho_{_{C}}} = \left ( \frac{8 \pi G}{3 H_0^2}
\right )\,\rho_M^0 $ and $ \ds
\Omega_{\Lambda}\,=\left(\frac{\Lambda}{8 \pi G}
\right)\frac{1}{\rho_{_{C}}}\,=\,\frac{\Lambda}{3\,H_0^2}$. Here,
$\rho_M^0$ is the actual value of the nonrelativistic matter
density and $H_0$, the actual value of the Hubble parameter.
$\Omega_{\Lambda}$ is the fraction of the critical energy density
contained in a smoothly distributed vacuum energy referred to as
$\Lambda$, and $\Omega_M$ represents the matter density related to
the baryonic and nonbaryonic CDM densities. The constant $G$ is
the Newton constant, and we use $c=1$ for the speed of light. From
now on, all quantities with upper or lower zero indexes specify
current values.

Due to measurements and theoretical arguments, it seems natural to
consider flat universe models, but one interesting question to ask
is whether this flatness is due to a sort of compensation among
different components that enter into the dynamical equations. In
this respect, our main goal in this paper is to address this sort
of question. In the literature we have found some descriptions
along these lines. For instance, closed models with an important
matter component with equation of state given by $P = - \rho / 3$
have been studied. Here, the universe expands at constant
speed~\cite{Ko}. Other authors, by using the same properties for
the universe, have added a nonrelativistic matter density in which
$\Omega_T$ is less than one, thus describing an open
universe~\cite{KaTo}. Also, flat decelerating models have been
simulated~\cite{CrdCHe}. The common fact in all of these models is
that, even if the starting geometry presents curvature, all models
are indistinguishable from flat geometries at low redshift.

In this paper we wish to consider universe models that have
curvature and are composed by two-matter components. One of these
components is the usual nonrelativistic dust matter; the other
corresponds to a sort of quintessence-type matter, described by a
scalar field that we designate by $Q$. This field is fundamental
in the sense that it is introduced from a Lagrangian. In order to
mimic a flat Quintessence CDM,  we introduce a new scalar field
$\chi$, such that this field is constrained by recent astronomical
data. Therefore, in a flat background the scalar field $\chi$
together with the CDM component form the basis for the $\chi$CDM
model, a model that is restricted by the present observations. In
this way, it is the scalar field $\chi$, not the $Q$-field, that
is related to the observable quantities~\cite{Wa-etal}. Thus, we
assume that the effective equation of state for $\chi$ is given by
$P_{\chi} = w_{_\chi} \rho_{\chi}$, where $w_{_\chi}$ is the
observable effective equation state parameter~\cite{CaDaSt}. We
note that the astronomical observations (related to the type Ia
supernovae measurements) put an upper limit on the present value
of this parameter, $w_{_\chi} < -1/3$~\cite{Ga-etal}.

We also assume that the scalar field $Q$ is characterized by a
similar equation of state \be P_{_Q} = w_{_Q} \rho_{_Q},
\label{es1} \en where the parameter $w_{_Q}$ is, in general, a
variable quantity. Therefore, our goal in this paper is to
investigate the conditions under which a scenario with positive
(or negative) curvature may mimic a flat universe at low
redshifts. This approach forces us to determine the exact
contribution of the scalar field $Q$, together with the curvature
term, that gives rise to the flat quintessence accelerating
$\chi$CDM universe. However, we should comment that absorbing the
curvature term in a redefinition of the $Q$ field is certainly not
equivalent to getting a really flat universe, since the curvature
is a geometrical property, which follows directly from the metric
tensor and which enters into the FRW line element. As we mentioned
above, these two models become indistinguishable only for low
redshifts and are similar to the cases studied in
ref.~\cite{KaTo,CrdCHe}.  We will return to this point later on.
Here, we emphasize that our approach rests in the fact that there
is a clear similitude between the curvature and the flat universe
models at low redshift.


\section{\label{sec:level2}The dynamic field equations }

We start with the following effective action
$$ \hspace{-4.0cm}
\ds S\,=\,\int{d^{4}x\,\sqrt{-g}}\,\left
[\,\frac{1}{16\pi\,G}\,R\,\right .
$$ \be \hspace{2.0cm} \ds \left
.+\,\frac{1}{2}\,(\partial_{\mu}Q)^2\, -\,V(Q)\,+\,L_{M} \right ],
\label{ac1} \en where $R$ is the scalar curvature, $V(Q)$ is the
scalar potential associated to the field $Q$ and $L_M$ is related
to any ordinary matter component.

We shall assume that the $Q$ field is homogeneous, i.e. it is a
time-depending quantity only, $Q = Q(t)$, and the spacetime is
isotropic and homogeneous, with metric corresponding to the
Friedman-Robertson-Walker metric
$$ \hspace{-3.0cm} \ds d{s}^{2}= d{t}^{2}- a(t)^{2}\, \left [
\frac{dr^2}{1-k r^2} \right . $$
\begin{equation}
\hspace{2.0cm} \left.  +\,r^2 \left(\,d\theta^2+ sin^2 \theta
\,d\phi^2 \right) \frac{}{}\,\right ], \label{me1}
\end{equation}
where $a(t)$ represents the scale factor and the  parameter $k$
takes the values $k\,=\,-\,1,\, 0,\, 1$ corresponding to an open,
flat, closed three-geometry, respectively.  With these
assumptions, the action (\ref{ac1}) yields the following field
equations. The time-time component of Einstein equation  \be \ds
H\,^{2}\,=\,\frac{8\pi\,G}{3}\, \left
(\,\rho_{_{M}}\,+\,\rho_{_{Q}}\,\right )\, -\,\frac{k}{a^{2}},
\label{h1} \en the evolution equation for the scalar field $Q$ \be
\ds \ddot{Q}\, +\,3\,H\,\dot{Q}\,=-
\,\frac{\partial{V(Q)}}{\partial{Q}}. \label{ddq} \en and the
energy conservation law for the ordinary matter \be \ds
\dot{\rho}_{_{M}}+ 3 H (\rho_{_{M}}+P_{_{M}})=0 \label{rhom}.\en
In these Equations the overdots denote derivatives with respect to
$t$, $ \ds H\,=\,\frac{\dot{a}}{a}$ defines the Hubble expansion
rate, $\rho_{_{M}}$ and $\rho_{_{Q}}$ are the effective matter
energy density and the average energy densities, respectively. The
Q-energy density is defined by \be \label{rq} \ds
\rho_{_{Q}}\,=\,\frac{1}{2}\dot{Q}^2\,+\,V(Q)\,. \en We introduce
its average pressure $P_{_Q}$, by means of \be \label{pq} \ds
P_{_{Q}}\,=\,\frac{1}{2}\dot{Q}^2\,-\,V(Q)\, . \en These two
latter quantities are related by the equation of state,
Eq.~(\ref{es1}). Thus, expression~(\ref{ddq}) becomes \be
 \dot{\rho}_{_{Q}}+ 3 H
(\rho_{_{Q}}+P_{_{Q}})=0, \label{erq1} \en which, similar to the
ordinary matter, represents an energy balance for the scalar field
$Q$.  From now on, we consider the ordinary matter to correspond
to dust, which becomes characterized by the equation of state
$P_{_M} = 0 $. For this case we could solve the energy
Equation~\ref{rhom} analytically, in which case we get $\rho_{_M}
\propto a^{-3}$.

Summarizing, we have a combination of two noninteracting perfect
fluids : one, a dust matter component, ($\rho_{_{M}}$); the other,
the scalar field ($\rho_{_Q}$) component.

Eq. (\ref{h1}) may be written as $$ \ds \hspace{-4.0cm} H^2
\,=\,H_0^2\,\left[ \Omega_{_M}\,\left
(\frac{\rho_{_M}}{\rho_{_M}^0}\right ) \,\right. $$ \be
\hspace{3.0cm} \left. +\,\Omega_{_Q}\,\left(
\frac{\rho_{_Q}}{\rho_{_Q}^0}\right)\,+\,\Omega_{_k}\,\left(
\,\frac{a_0}{a} \,\right) ^2 \right].\label{h11}\en  Here, the
actual curvature density, $\Omega_k$, and the quintessence
density, $\Omega_Q$, parameters are defined by $ \ds \Omega_k
\,=\,-\,k\,\left (\frac{1}{a_0\,H_0} \right )^2,\label{ok}$ and $
\label{oq} \ds \Omega_Q\, =\,\left (\,
\frac{8\,\pi\,G}{3\,H_0^2}\, \right )\,\rho_{_Q}^0$, respectively.

In the next section we study the model that arises when the matter
component, $\rho_{_{M}}$ together with
Eqs.~(\ref{h1}),~(\ref{ddq}) and the equation of state for the
field $Q$, complement in such a way that a flat CDM accelerated
universe originates, specifically, the $\chi$CDM model.

In order to  simulate a flat universe, we assume that the energy
density  $\rho_{_{Q}}$ and the curvature term combine so that \be
\ds \frac{8 \pi G}{3}\rho_{_{Q}}(t) -\frac{k}{a^2(t)} =\frac{8 \pi
G}{3}\rho_{_{\chi}}(t), \label{c1} \en or equivalently, \be \ds
\Omega_{_Q}\,\left(
\frac{\rho_{_Q}}{\rho_{_Q}^0}\right)\,+\,\Omega_{_k}\,\left(
\,\frac{a_0}{a} \,\right) ^2
=\Omega_{\chi}\,\left(\,\frac{\rho_{_\chi}}{\rho_{_\chi}^0}\,\right)
 \label{c2}, \en where, similar
to the definitions of $\Omega_{_M}$ and $\Omega_{_Q}$, we have
defined $\ds \Omega_{\chi}\, =\,\left (\,
\frac{8\,\pi\,G}{3\,H_0^2}\, \right )\,\rho_{_{\chi}}^0$. This
latter quantity is related to the recent astronomical measurement
of distant supernova of type Ia.

Note that we could get an explicit expression (as a function of
time) for the unknown energy density, $\rho_{_\chi}$, if we know
both the scale factor $a(t)$ and the time dependence of the
$Q$-density, $\rho_{_Q}$. Note also that for $\Omega_k > 0$ (open
universes) we must have $ \ds \rho{_{\chi}} >
\frac{|\Omega_k|}{\Omega_\chi}\rho_{_{\chi}}^0\left(\frac{a_0}{a}\right)^2
$, since $\rho_{_{Q}} > 0$. At this point we should comment on the
difference between the two scalar fields $Q$ and $\chi$. As we
saw, the scalar field $Q$ is defined from the fundamental
Lagrangian, from which we could define the stress-energy density
tensor with the property of a perfect fluid behaviour, which
allows us to introduce the pressure $P_Q$. However, we could not
say the same of $\chi$. The definition of the stress-energy
density tensor associated with the $\chi$ field is more subtle, as
we can see in Eq.~(\ref{c1}) (or~(\ref{c2})). We hope to address this
problem in subsequent work. Here, we just take $P_\chi$ as an
effective pressure that follows the equation of state $P_\chi =
w_{_{\chi}} \rho_{_{\chi}}$, where $w_{_{\chi}}$ is determined
from the observational data.



\section{Characteristics of the model}

In this section we will impose explicit conditions under which a
curved universe ($k = \pm 1$) may look similar to a flat universe
($k = 0$) at low redshift. This flat model is defined by
expression (\ref{c1}) (or (\ref{c2})), which reduces the time-time
Einstein Eq.~(\ref{h11}) to \be \ds H\,^{2}\,=\,H_0^2\,\left [ \,
\Omega_{M}\,\left( \frac{ \rho_{_M}}{\rho_{_M}^0}\right) +\,
\Omega_{\chi}\,\left( \frac{ \rho_{_\chi}}{\rho_{_\chi}^0}\right)
\,\right ]. \label{h4}\en

Eqs.~(\ref{c2}) and~(\ref{h4}), along with the evolution equations
for the scalar fields $\chi$ and $Q$, form the basic set of
equations that describes our model.

Note that, when Eq.~(\ref{c2}) is evaluated at present time, we
obtain the following relation among the omega parameters \be \ds
\Omega_Q\,=\,\Omega_{\chi} \,- \,\Omega_k.\label{qkx}\en From this
relation we observe that, for $k=1$, $\Omega_Q$ must be greater
than $\Omega_{\chi}$, since $\Omega_{k=1} < 0$. For $k = -1$,
$\Omega_{k} > 0$, thus $\Omega_{\chi} > \Omega_Q$. Notice also
that Eq.~(\ref{h4}) gives the additional expression $\Omega_M +
\Omega_\chi \equiv \Omega_T = 1$ when evaluated at present time.
From the present observational values of $\Omega_{\chi}$ and
$\Omega_k$, we could get the actual value of the parameter
$\Omega_Q$.   Note also that when Eqs.~(\ref{h11})  and~(\ref{h4})
are compared we observe that the variables $\rho_{_{M}}$ and $H$
(and also the scale factor $a$, as we can see from Eq.~(\ref{c1})
or equivalently Eq.~(\ref{c2})) appear {\it identical} in the two
scenarios.

In order to see that a curvature model at low redshift is
indistinguishable from a flat one, we could consider the
luminosity distance $D_L$ as a function of the redshift $z$ or the
angular distance~\cite{CrdCHe}. Let us take the first one. The
luminosity distance between a source at a redshift $z>0$ and
$z=0$, related to a curvature model is obtained from the
expression \be \ds D_L^{k \neq 0}(z)\, \propto \,\frac{
(1+z)}{\sqrt{\mid \Omega_k\mid}}sin\left [\sqrt{\mid
\Omega_k\mid}\xi(z) \right]. \en   For a flat universe model we
obtain \be \ds D_L^{k = 0}(z)\, \propto \, (1+z)\xi(z). \en In
these expressions we have defined   $\xi(z)$ by means of $ \ds
\xi(z)= \frac{1}{a_0}\int_0^z\frac{dz}{H(z)}$ which represents the
polar-coordinated distance between a source at $z$ and another at
$z=0$ in the same line of sight. For $z \ll 1$ we could show that
$\xi(z)\propto z$. Thus, since $\mid \Omega_k\mid \ll 1$ we
observe that the luminosity distance in both cases coincide.
Therefore, we expect that the differences between the curvature and
the flat models happen to high enough redshifts.

When the quintessence component has a constant equation state
parameter,  $w_{_{\chi}}^0 \equiv w_{_{\chi}}$, which  is a
negative number that lies in the range $-1< w_{_{\chi}} < - 0.3$,
we could immediately solve for the density $\rho_{_{\chi}}$ as a
function of the scale factor \be \ds \rho_{_{\chi}}(a)=
\rho_{_{C}}\,\Omega_{\chi} \left(
\frac{a_0}{a}\right)^{3(1+w_{_{\chi}})}.\label{rx}\en Thus,
Eq.~(\ref{h4}) becomes \be \ds H(a)\, =\, H_0\,\left(\frac{a_0}{a}
\right)^{3/2}\,\sqrt{\Omega_M\,+\,\Omega_{\chi}\left(
\frac{a_0}{a}\right)^{3 w_{_{\chi}}}}. \label{h44} \en The
solution of Eq.~(\ref{h44}) is given by
$$ \hspace{-4.0cm}\ds  t\,=\,
\frac{2}{3\,H_0\,\sqrt{\Omega_M}}\,\left(
\frac{a}{a_0}\right)^{3/2}\,
$$
\be  \ds \times\,_2F_1\left(\frac{1}{2}, -\,\frac{1}{2 w_{_{\chi}}
}; 1- \frac{1}{2  w_{_{\chi}} }; -\left(\frac{\Omega_k +
\Omega_Q}{\Omega_M}\right)\,\left(\frac{a}{a_0} \right)^{-
3\,w_{_{\chi}}} \right),\label{ta1}\en where $_2F_1$ represents
the generalized hypergeometric function. The initial condition
that we have used in solving Eq.~(\ref{h44}) is $a = 0$ at $t =
0$.

We use the definition of  $P_{_{\chi}}$ and $\rho_{_{\chi}}$ in
terms of the scalar field $\chi$, together with the equation of
state that relates these quantities, for obtaining $\chi$ as a
function of the scale factor. The result is
$$ \ds \hspace{-5.0cm}\chi(a)\,=\,\chi_{_{0}}\left(\frac{a}{a_0}\right)^{-3w_\chi/2} $$
\be
\ds \times \left[\,\frac{_2F_1\left(\frac{1}{2}, \frac{1}{2};
\frac{3}{2};-\left(\frac{\Omega_k +
\Omega_Q}{\Omega_M}\right)\,\left(\frac{a}{a_0} \right)^{- 3\,
w_{_{\chi}}} \right)}{_2F_1\left(\frac{1}{2}, \frac{1}{2};
\frac{3}{2};-\left(\frac{\Omega_k + \Omega_Q}{\Omega_M}\right)
\right)}\right], \label{chia} \en where $\chi_{_{0}}$ is given by
$\ds \chi_{_{0}}=\widetilde{\chi}\,\,_2F_1\left(\frac{1}{2},
\frac{1}{2}; \frac{3}{2};-\left(\frac{\Omega_k +
\Omega_Q}{\Omega_M}\right) \right)$, with $\widetilde{\chi}=
\sqrt{4\rho_c/9H_0^2}\sqrt{(\Omega_k +
\Omega_Q)(1+w_\chi)/\Omega_M w_\chi^2}$.

In a similar way, we get for the scalar potential $V_{_{\chi}}$
\be \ds V_{_{\chi}}(a)\,=\,V_\chi^0\,\left( \frac{a_0}{a}\right)
^{3(1+w_{_{\chi}})}, \label{vchi} \en where $\ds
V_\chi^0\,=\,\frac{1}{2}(1- w_{_{\chi}})\,\rho_{_{C}}(\Omega_k +
\omega_Q)$.

\begin{figure}[ht]
\includegraphics[width=3.0in,angle=0,clip=true]{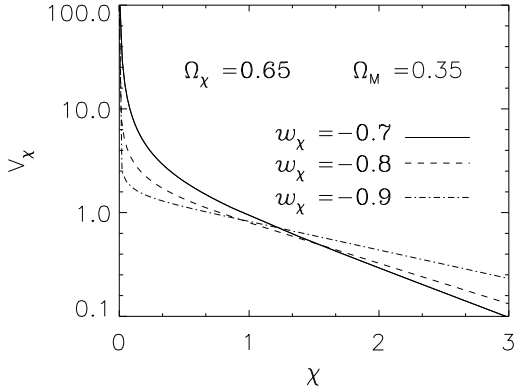}
\caption{ This graph shows the scalar potential $V_{_{\chi}}$ as a
function of the scalar field $\chi$ for $w_{_{\chi}}=-0.7$, $
-0.8$ and $ -0.9 $. We use $\Omega_M=0.35$ and $\Omega_\chi
=0.65$.} \label{V(chi)}
\end{figure}

Fig~(\ref{V(chi)}) shows the plot of the scalar potential
$V_{_{\chi}}$ as a function of $\chi$, for three different values
of the equation state parameter $w_{_{\chi}}$. The parameters
$\Omega_\chi$ ($= \Omega_k + \Omega_Q$) and $\Omega_M$ have been
fixed at values 0.65 and 0.35, respectively. This form of
potential has been described in the literature~\cite{DeRaSaSt}.
Note that, in the limit $w_{_{\chi}} \longrightarrow -1$ the
potential $V_{\chi} \longrightarrow const. \equiv
\rho_{_C}\,\Omega_{\chi}$, i.e., the model becomes equivalent to
the $\Lambda$ Cold Dark Matter scenario\cite{CaDaSt}.

During the evolution of this model, we distinguish two periods:
one, where the regular matter dominates ($\rho_{_M} \gg
\rho_{_{\chi}}$); the other, where $\rho_{_{\chi}} \gg \rho_{_M}$.
It is not hard to see that the time at which $\rho_{_M}$ becomes
equal to $\rho_{_{\chi}}$ is given by
$$\hspace{-3.0cm} \ds t_{eq}\,=\,\frac{2}{3\,H_0\,\sqrt{\Omega_M}}\left(\frac{\Omega_M}
{\Omega_{\chi}}\right)^{-\frac{1}{2 w_{_{\chi}}}}$$ \be \ds \times
_2F_1\left(\frac{1}{2},-\frac{1}{2 w_{_{\chi}}};1-\frac{1}{2
w_{_{\chi}}}; -1\right).\label{teq2} \en

Notice that the age of the universe in this model is given by
 $$ \hspace{-3cm}
 \ds t_0= \frac{2}{3 H_0 \sqrt{\Omega_M}} $$ \be \ds
 \times \,_2F_1\left(\frac{1}{2},- \frac{1}{2 w_{_\chi}}
 ; 1-\frac{1}{2 w_{_\chi}}; -\frac{\Omega_\chi}{\Omega_M}\right),
 \label{tchi}
 \en
 \begin{figure}[ht]
\includegraphics[width=3.0in,angle=0,clip=true]{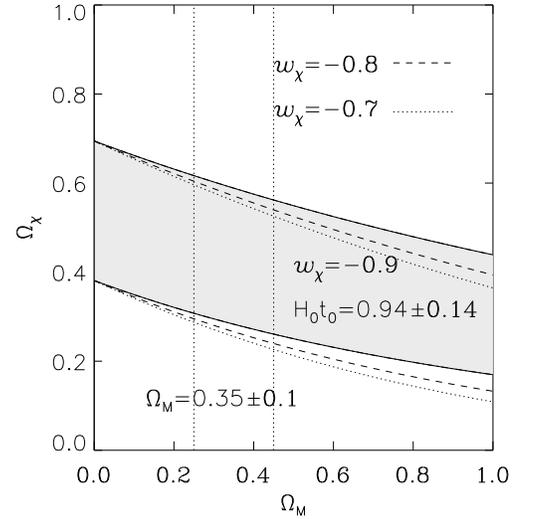} \caption{ This graph
represents $\Omega_\chi$ as a function of $\Omega_M$ for the range
of $t_0\,H_0= 0.94 \pm 0.14$ for three different values of the
equation state parameter, $w_{_{\chi}}= -0.7, -0.8, -0.9 $. The
dark region corresponds to  $w_{_{\chi}}=-0.9$. The vertical
dotted lines show the observational range for the $\Omega_M$
parameter.} \label{age}
\end{figure}
where, similar to $t_{eq}$, depends on the observable parameters
$\Omega_\chi$, $\Omega_M$ and $w_{_\chi}$. For $\Omega_M = 0.3$,
$\Omega_\chi = 0.7$ and $w_{_\chi} = -0.8$ this gives $t_0\,H_0
\sim 0.93$, which lies in the observational range $t_0\,H_0 = 0.94
\pm 0.14$ of the measurement of the age of the
universe~\cite{MST}. Fig~(\ref{age}) shows $\Omega_\chi$ as a
function of $\Omega_M$ for the range of $t_0\,H_0$ specified
above. The dark region represents the value $w_{_{\chi}}=-0.9$.

\begin{figure}[ht]
\includegraphics[width=3.in,angle=0,clip=true]{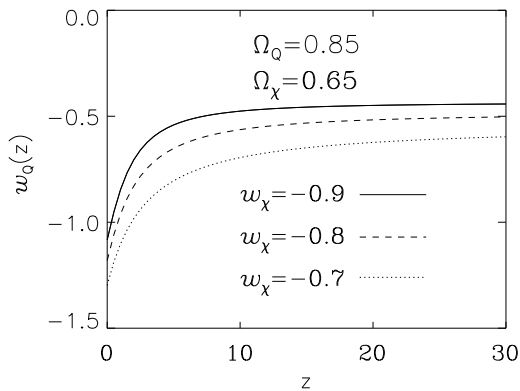} \caption{
The graph shows    $w_Q$ (in unit of $|w_{_{Q}}^0|$) as a function
of the redshift $z$ for three different values of the equation
state parameter, $w_{_{\chi}}= -0.7, -0.8, -0.9 $. We have used
the values  $\Omega_Q=0.85$ and $\Omega_\chi=0.65$, corresponding
to a closed model.} \label{wz_QCDM}
\end{figure}

We are now going to describe the properties of the scalar field
$Q$. Following an approach analogous to that used for the scalar
field $\chi$, we find that the parameter $w_Q$, is given by
$$ \hspace{-2.0cm}w_Q(a)=-\mid w_Q^0 \mid \left(
\frac{1+\beta}{1-3\beta\, w_{_{\chi}}}\right) $$
 \be \ds \hspace{1.0cm}\times \left[ \frac{1-3\beta\,
w_{_{\chi}} \left( \frac{a}{a_0}\right)^{-3 w_{_{\chi}}
-1}}{1+\beta\left( \frac{a}{a_0}\right)^{-3
w_{_{\chi}}-1}}\right],\en where $\beta = \Omega_\chi/\mid
\Omega_k \mid$ and $w_Q^0$ is the actual value of $w_Q$ defined by
$w_Q^0=P_Q^0/\rho_{_{Q}}^0$. Fig (\ref{wz_QCDM}) shows its
dependence with the  redshift $z$ defined as $z\equiv a_0/a-1$,
for three different values of the $w_{_{\chi}}$ parameter. For
completeness, we have chosen $\Omega_Q=0.85$ and
$\Omega_\chi=0.65$ in this plot, i.e. we have considered a closed
model.

The scalar field $Q$ results in  \be \ds
Q(a)=\overline{Q}\int_0^{a/a_0}\sqrt{
\frac{1+\frac{3}{2}\beta(1+w_\chi)x^{-(1+3w_\chi)}}{x+\frac{\Omega_\chi}{\Omega_M}
x^{-(1+3w_\chi)}}} dx,\label{Q_QCDM}\en where
$\overline{Q}=\sqrt{\frac{1}{4 \pi G}\left(\frac{\mid \Omega_k
\mid}{\Omega_M} \right)}$.

\begin{figure}[ht]
\includegraphics[width=3.0in,angle=0,clip=true]{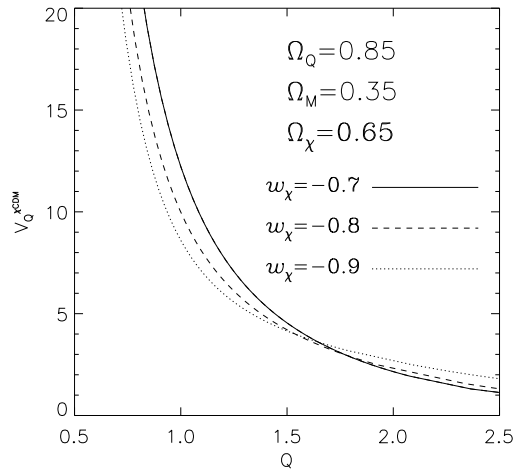} \caption{
The graph shows $V_Q$ (in unit of $\rho_{_C}/2$) as a function of
$Q$ (in unit of $1/\sqrt{4 \pi G}$) for three different values
$w_{_{\chi}}= -0.7, -0.8$, and $-0.9$. We use the values
$\Omega_Q=0.85$, $\Omega_\chi=0.65$ and $\Omega_M=0.35$ associated
with a closed model.} \label{potencial2_QCDM}
\end{figure}

The scalar potential, $V_Q(a)$, is found to be given by
$$\ds \hspace{-5.0cm}V_Q(a) = V_Q^0\left(\frac{a_0}{a}\right)^2$$
\be \ds \hspace{2.0cm} \times \left[
\frac{4+3\beta(1-w_\chi)\left(\frac{a}{a_0}\right)^{-(1+3w_\chi)}}{4+3\beta(1-w_\chi)}
\right], \label{Vq_QCDM}\en where $V_Q^0 = \frac{\rho_C
\Omega_Q}{2(1-w_\chi)}\left[4/3 + \beta(1-w_\chi)\right]$. Figure
(\ref{potencial2_QCDM}) shows the potential $V_Q$ as a function of
the scalar field $Q$ for three different values of $w_\chi$.
Again, we have considered a closed model with $\Omega_Q=0.85$,
$\Omega_M=0.35$ and $\Omega_\chi=0.65$. For open models, i.e. for
$\Omega_\chi > \Omega_Q$ these curves are very similar.

Note that the potential decreases when $Q$ increases. We observe
that this potential asymptotically tends to vanish for $a
\longrightarrow \infty $. This implies that, asymptotically, the
effective equation of state becomes  $P_Q = \rho_{_{Q}}$ (for
$\dot{Q}\,\neq 0$), corresponding to a stiff fluid.

\section{The quintessence scalar potential}

One of the possible modifications to the basic idea of
quintessence includes  the use of other
potentials~\cite{AACS,PBJM,IZPS}. It may be interesting to
investigate the situation in which the scalar potential $V(\chi)$
is of the form~\cite{YG} \be \ds V(\chi) = \alpha \dot{\chi}^2 +
\gamma \label{v},\en where $\alpha$ and $\gamma$ are two
constants. We may write $\gamma$ as $\gamma = (\alpha_{_{-}} -
\alpha_{_{+}}\,w_{_{\chi}})\rho_{_{\chi}}^0$, where
$\alpha_{_{\pm}} = \frac{1}{2} \pm \alpha$.

Combining Eq.~(\ref{v}) with the evolution Equation for the field
$\chi$ we get $\dot{\chi}$ as a function of the scale factor $a$,
and thus we get $$\hspace{-3.0cm} \ds \rho_{_{\chi}}(a) =
\rho_{_{\chi}}^0\left [\alpha_{_{+}}\left(\frac{a_0}{a}
\right)^{3/\alpha_{_{+}}} + \alpha_{_{-}}\right ] $$ \be
\hspace{3.7cm} + \alpha_{_{+}} P_{_{\chi}}^0\left
[\left(\frac{a_0}{a} \right)^{3/\alpha_{_{+}}} - 1\right ],
\label{d3} \en and
$$\hspace{-3.0cm} \ds P_{_{\chi}}(a) = P_{_{\chi}}^0\left
[\alpha_{_{-}}\left(\frac{a_0}{a} \right)^{3/\alpha_{_{+}}} +
\alpha_{_{+}}\right ] $$
  \be \hspace{3.7cm} + \alpha_{_{-}} \rho_{_{\chi}}^0\left
[\left(\frac{a_0}{a} \right)^{3/\alpha_{_{+}}} - 1\right ],
\label{p3} \en where the constants $\rho_{_{\chi}}^0$ and
$P_{_{\chi}}^0$ represent the actual energy density and pressure,
respectively, and are related to an arbitrary integration
constant. From these two latter relations we get \be \ds
w_{_{\chi}}(a)=\frac{w_{_{\chi}}\left
[\alpha_{_{-}}\left(\frac{a_0}{a}
\right)^{\frac{3}{\alpha_{_{+}}}} + \alpha_{_{+}}\right ]+
\alpha_{_{-}}\left [\left(\frac{a_0}{a}
\right)^{\frac{3}{\alpha_{_{+}}}} - 1\right ]}{\left
[\alpha_{_{+}}\left(\frac{a_0}{a}
\right)^{\frac{3}{\alpha_{_{+}}}} + \alpha_{_{-}}\right ]+
w_{_{\chi}} \alpha_{_{+}}\left [\left(\frac{a_0}{a}
\right)^{\frac{3}{\alpha_{_{+}}}} - 1\right ]},\label{w3}\en
\\where, just as before, we take $\ds w_{_{\chi}} =
\frac{P_{_{\chi}}^0}{\rho_{_{\chi}}^0}$. Note that for $\ds
w_{_{\chi}} =\frac{\alpha_{_{-}}}{\alpha_{_{+}}}$ we get
$w_{_{\chi}}(a)=w_{_{\chi}} = const.$ This case corresponds to use
$\gamma = 0$ in Eq.~(\ref{v}). As we will soon see, this case
allows us to write down an explicit expression for the
quintessence scalar potential, $V(\chi)$. Through this paper, we
will address this case only.

We could solve the time-time Einstein Equation~(\ref{h4}) to
obtain in this case
$$\hspace{-4.0cm}  \ds t=\frac{1}{H_0}\frac{2}{\sqrt{\Omega_M}} \left(
\frac{a}{a_0}\right)^{1/2}\,$$ \be \hspace{1.3cm} \ds \times\,\,
_2\,F_1\left(\frac{1}{2},
\frac{\alpha_{_{+}}}{\alpha_{_{-}}};\frac{5}{6};-\frac{\Omega_\chi}
{\Omega_M}\left(\frac{a}{a_0}\right)^{3\,\alpha_{_{-}}/\alpha_{_{+}}}
\right).\label{t3} \en

The quintessence scalar field $\chi$ in terms of the scale factor
is given by
$$ \hspace{-2.0cm}\ds \chi(a) = \chi_{_{0}}+ \varepsilon\,\, arcsinh\left[ \delta
\left(\frac{a}{a_0}\sqrt{1+\delta^2}\right. \right. $$ \be
\hspace{4.0cm} \ds \left.\left. -\sqrt{1+\delta^2 \left
(\frac{a}{a_0} \right)^2} \right)\right], \label{f3} \en where
$\ds \varepsilon =\frac{2}{3 H_0 \mid
w_{_{\chi}}\mid}\sqrt{\frac{2
V_{\chi}^0}{\Omega_{\chi}}\,\frac{1}{\alpha_{_{-}}+\,\alpha_{_{+}}}}$
and $\ds \delta = \left(
\frac{\Omega_\chi}{\Omega_M}\right)^{1/3|w_{_{\chi}}|}$.

Eq.~(\ref{f3}) allows us to write down an explicit expression for
the dark energy scalar potential $V(\chi)$. We get \be \ds V(\chi)
= V_\chi ^0 \left\{ \frac{\delta} {sinh\left[ \varepsilon^{-1}
\left(\chi - \chi_0\right) + arcsinh(\delta)\right]
}\right\}^{3/\alpha_{_{+}}},\label{v2} \en where $V_\chi^0$
represents the actual value of this potential. The hyperbolic
potential~(\ref{v2}) has been used in the
literature~\cite{YG,TMLU,VJ}. This potential was studied for
getting tracker solution from the corresponding field equations.

One of the characteristics of the scalar field $Q$ is given by the
equation state parameter $w_{_Q}(a)$, which is given by
\begin{widetext} \be \ds w_{_Q}(a) =\frac{(3
w_{_Q}^0+1)(w_{_\chi}+1)\alpha_{_{-}}\left(\frac{a_0}{a}\right)^{3/\alpha_{_{+}}}\,
+\,(w_{_Q}^0\,-\,w_{_{\chi}})\left(\frac{a_0}{a}\right)^2}{(3
w_{_Q}^0+1)(w_{_\chi}+1)\alpha_{_{-}}\left(\frac{a_0}{a}\right)^{3/\alpha_{_{+}}}\,
-\,(w_{_Q}^0\,+\,w_{_{\chi}})\left(\frac{a_0}{a}\right)^2},.\label{wq3}\en
\end{widetext}
Here, the quantities $w_{_{Q}}^0$ are defined by \be \ds
w_{_{Q}}^0 =
\frac{w_{_{\chi}}\,\Omega_\chi\,+\,\Omega_k/3}{\Omega_\chi\,-\,\Omega_k}
\,=\,\left(w_{_{\chi}} + \frac{1}{3}
\right)\frac{\Omega_\chi}{\Omega_Q}\,-\,\frac{1}{3}\label{w0}, \en
where we have used the relation $\Omega_k \,+\,\Omega_Q
\,=\,\Omega_\chi$ in the latter expression.

With  $x \equiv a/a_0$, the scalar potential associated to the $Q$
field is given by
$$ \ds V_Q(x)\,=\,\rho_{_Q}^0
\left[\left(\frac{\alpha_{_{+}}\,+\,\alpha_{_{-}}}{2}\right)\left(1\,+\,w_{_{\chi}}
\right)x^{-3/\alpha_{_{+}}}\,\right.$$ \be \hspace{3.0cm} \ds
\left. -\,\frac{2}{3}\frac{\Omega_k}{\Omega_M}x^{-2}\, \right]
\label{vq2}, \en and the corresponding scalar field $Q$ is
expressed by means of the following integral $$ \hspace{-5.0cm}
\ds Q(x)\,=\,\frac{\sqrt{\rho_{_Q}^0}}{H_0}\int_x ^1\,\frac{dz}{z}
$$ \be \ds {\times
\sqrt{\frac{(1+w_{_{\chi}})\frac{\Omega_\chi}{\Omega_Q}\,
z^{-3\alpha_{_{-}}/\alpha_{_{+}}}-\frac{2}{3}\frac{\Omega_k}{\Omega_Q}z}{1\,+\,(1+
w_{_{\chi}})\frac{\Omega_\chi}{\Omega_M}\,\alpha_{_{+}}\,
z^{-3\alpha_{_{-}}/\alpha_{_{+}}}}}} .\label{q3} \en A numerical
integration shows that this potential presents the same
characteristic than that described in the previous case, i.e.
$V(Q)$ decreases when $Q$ increases, reaching the limit
$V(Q)\longrightarrow 0$ for $Q \longrightarrow \infty$.

Finally, one interesting parameter to determine in this kind of
model is the deceleration parameter, which is defined by $\ds q =
-\frac{\ddot{a}}{\dot{a} H^2}$, and when evaluated to present
time, it becomes \be \ds q_{_{0}} = \frac{1}{2} \Omega_M -
\frac{1}{2}\left( 3 |w_{_{\chi}}|-1 \right)\left( \Omega_k +
\Omega_Q \right). \label{q0} \en  For $w_{_{\chi}} < -1/3(1+\Omega_M/\Omega_\chi)$,
the parameter $q_{_{0}}$ becomes negative, in agreement with the observed
acceleration of the universe.

\section{conclusions}

In this paper we have described a curvature universe model in
which, apart from the usual CDM component, we have included a
quintessence-like scalar field $Q$. We have fine-tuned the energy
density, associated with this field, together with the curvature,
for mimicking  a flat model which resembles the quintessence (or
dark energy) $\chi CDM$ model, which is characterized by
$\Omega_T= \Omega_M+\Omega_\chi= 1$. We have assumed for $Q$ an
effective equation of state $P_Q =w_Q \rho_{_{Q}}$, with $w_Q <
0$. We have determined the form of the potential associated with
this field, which effectively has the property of a quintessence
potential, i.e. $V(Q)$ decreases when $Q$ increases, approaching
zero asymptotically. Under the assumption that the equation state
parameter was constant, we could describe explicitly the
characteristics of the dark energy scalar field, $\chi$.
Certainly, the basic idea of quintessence rests on the
determination of the scalar potential as a function of this field.

Under an appropriate  choice for the dark energy scalar potential,
$V(\chi)$ (in terms of $\dot{\chi}$), we determine explicit
expression for this potential as a function of $\chi$. In this
case we also gave expression for the scalar field $Q$ and its
potential. Here, we determine the deceleration parameter which,
under an appropriate choice of the parameters that characterize
the model, became positive, in agreement with the acceleration
detected by astronomical observations.

\begin{acknowledgments}
I am grateful to Francisco Vera for plotting assistance.
Discussion with M. Cataldo, N. Cruz, S. Lepe, F. Pe\~na and P.
Salgado are acknowledged. I also acknowledge the Universidad de
Concepci\'on and Universidad del Bio-Bio for partial support of
the Dichato Cosmological Meeting, where part of this work was
done. This work was supported from COMISION NACIONAL DE CIENCIAS Y
TECNOLOGIA through FONDECYT Projects N$^0$ 1000305 and 1010485
grants, and from UCV-DGIP N$^0$ 123.752.
\end{acknowledgments}

\end{document}